\documentclass[a4paper,fleqn]{cas-dc}

\usepackage[numbers]{natbib}
\usepackage{upgreek} 
\usepackage{ulem}
\usepackage[dvipsnames]{xcolor}

\def\tsc#1{\csdef{#1}{\textsc{\lowercase{#1}}\xspace}}
\tsc{WGM}
\tsc{QE}
\tsc{EP}
\tsc{PMS}
\tsc{BEC}
\tsc{DE}

\begin{document}
\let\WriteBookmarks\relax
\def\floatpagepagefraction{1}
\def\textpagefraction{.001}
\shorttitle{}
\shortauthors{A. Pandey et~al.}

\title [mode = title]{Timing performance of large prototype based on $\upmu$RWELL- PICOSEC detector technology with $10 \times 10\ \mathrm{cm}^{2}$ active area  }                      

\tnotetext[1]{The research described in this article was conducted under the Laboratory Directed Research and Development (LDRD) Program at Thomas Jefferson National Accelerator Facility for the U.S. Department of Energy under contract DE-AC05-06OR23177.}

\author[9]{A. Pandey}[type=editor,
                        auid=000,bioid=1,
                        prefix=,
                        role=
                     ]
\cormark[1]
\fnmark[1]
\ead{apandey@jlab.org}

\author[9]{K. Gnanvo}
\author[9]{B. Kross}
\author[9]{J. McKisson}
\author[9]{A. Weisenberger}
\author[9]{W. Xi}
\author[8]{J. Dutta}
\author[8]{N. Shankman}

\author[1]{L. Scharenberg}
\author[1]{J. Alozy}
\author[2,3]{Y. Angelis}
\author[4]{S. Aune}
\author[1]{R. Ballabriga}
\author[5]{J. Bortfeldt}
\author[1]{F. Brunbauer}
\author[6,7]{M. Brunoldi}
\author[1]{M. Campbell}
\author[1]{R. De Oliveira}
\author[10]{G. Fanourakis}
\author[1,11]{J.M. Fernandez-Tenllado}
\author[1,12]{K.J. Flöthner}
\author[6,7]{D. Fiorina}
\author[13]{M. Gallinaro}
\author[14]{F. Garcia}
\author[4]{I. Giomataris}
\author[11,15]{S. Gomez}
\author[4]{F.J. Iguaz}
\author[1]{D. Janssens}
\author[4]{A. Kallitsopoulou}
\author[16]{M. Kovacic}
\author[4]{P. Legou}
\author[1,17]{M. Lisowska}
\author[18]{J. Liu}
\author[12,19]{M. Lupberger}
\author[11]{R. Manera}
\author[1,2]{I. Maniatis}
\author[11]{A. Mariscal}
\author[11]{J. Mauricio}
\author[18]{Y. Meng}
\author[12,1]{H. Muller}
\author[1]{E. Oliveri}
\author[1,20]{G. Orlandini}
\author[4]{T. Papaevangelou}
\author[11]{E. Picatoste}
\author[1,21]{M. Piller}
\author[22]{M. Pomorski}
\author[1]{L. Ropelewski}
\author[2,3]{D. Sampsonidis}
\author[11]{A. Sanuy}
\author[1]{T. Schneider}
\author[22]{E. Scorsone}
\author[4]{L. Sohl}
\author[1]{M. van Stenis}
\author[23]{Y. Tsipolitis}
\author[2,3]{S.E. Tzamarias}
\author[24]{A. Utrobicic}
\author[6,7]{I. Vai}
\author[1]{R. Veenhof}
\author[6,7]{P. Vitulo}
\author[18]{X. Wang}
\author[25]{S. White}
\author[18]{Z. Zhang}
\author[18]{Y. Zhou}

\affiliation[1]{organization={European Organization for Nuclear Research (CERN)}, addressline={1211 Geneva 23}, country={Switzerland}}
\affiliation[2]{organization={Department of Physics, Aristotle University of Thessaloniki}, addressline={University Campus, 54124 Thessaloniki}, country={Greece}}
\affiliation[3]{organization={Center for Interdisciplinary Research and Innovation (CIRI-AUTH)}, addressline={Thessaloniki 57001}, country={Greece}}
\affiliation[4]{organization={Institut de Recherche sur les lois Fondamentales de l’Univers (IRFU, CEA)}, addressline={Université Paris-Saclay, F-91191 Gif-sur-Yvette}, country={France}}
\affiliation[5]{organization={Department for Medical Physics, Ludwig Maximilian University of Munich}, addressline={Am Coulombwall 1, 85748 Garching}, country={Germany}}
\affiliation[6]{organization={Dipartimento di Fisica, Università di Pavia}, addressline={Via Bassi 6, 27100 Pavia}, country={Italy}}
\affiliation[7]{organization={INFN Sezione di Pavia}, addressline={Via Bassi 6, 27100 Pavia}, country={Italy}}
\affiliation[8]{organization={Department of Physics and Astronomy, Stony Brook University}, addressline={Stony Brook, NY 11794-3800}, country={USA}}
\affiliation[9]{organization={Thomas Jefferson National Accelerator Facility (Jefferson Lab)}, addressline={12000 Jefferson Avenue, Newport News, VA 23606}, country={USA}}
\affiliation[10]{organization={Institute of Nuclear and Particle Physics, NCSR "Demokritos"}, addressline={P.O. Box 60037, 15310 Agia Paraskevi}, country={Greece}}
\affiliation[11]{organization={Institute of Cosmos Sciences (ICCUB), University of Barcelona}, addressline={Martí i Franquès 1, 08028 Barcelona}, country={Spain}}
\affiliation[12]{organization={Helmholtz-Institut für Strahlen- und Kernphysik, University of Bonn}, addressline={Nußallee 14-16, 53115 Bonn}, country={Germany}}
\affiliation[13]{organization={Laboratório de Instrumentação e Física Experimental de Partículas (LIP)}, addressline={Av. Prof. Gama Pinto 2, 1649-003 Lisbon}, country={Portugal}}
\affiliation[14]{organization={Helsinki Institute of Physics, University of Helsinki}, addressline={P.O. Box 64, FI-00014}, country={Finland}}
\affiliation[15]{organization={Electronics Department, Polytechnic University of Catalonia (UPC)}, addressline={Eduard Maristany 16, 08019 Barcelona}, country={Spain}}
\affiliation[16]{organization={University of Zagreb, Faculty of Electrical Engineering and Computing}, addressline={Unska 3, 10000 Zagreb}, country={Croatia}}
\affiliation[17]{organization={Université Paris-Saclay}, addressline={F-91191 Gif-sur-Yvette}, country={France}}
\affiliation[18]{organization={State Key Laboratory of Particle Detection and Electronics, University of Science and Technology of China (USTC)}, addressline={Hefei 230026}, country={China}}
\affiliation[19]{organization={Physikalisches Institut, University of Bonn}, addressline={Nußallee 12, 53115 Bonn}, country={Germany}}
\affiliation[20]{organization={Friedrich-Alexander-Universität Erlangen-Nürnberg}, addressline={Schloßplatz 4, 91054 Erlangen}, country={Germany}}
\affiliation[21]{organization={Institute of Electronics, Graz University of Technology}, addressline={Inffeldgasse 12/I, 8010 Graz}, country={Austria}}
\affiliation[22]{organization={Laboratory for Integration of Systems and Technology (CEA-LIST), Diamond Sensors Laboratory, CEA Saclay}, addressline={F-91191 Gif-sur-Yvette}, country={France}}
\affiliation[23]{organization={National Technical University of Athens}, addressline={106 82 Athens}, country={Greece}}
\affiliation[24]{organization={Ruder Bošković Institute}, addressline={Bijenička cesta 54, 10000 Zagreb}, country={Croatia}}
\affiliation[25]{organization={Department of Physics, University of Virginia}, addressline={P.O. Box 400714, Charlottesville, VA 22904-4714}, country={USA}}

\cortext[cor1]{Corresponding author}               

\begin{abstract}
The $\upmu$RWELL-PICOSEC detector, which combines a $\upmu$RWELL gaseous amplification structure with a Cherenkov radiator and photocathode, is a novel approach to acheive fast and precise timing in gaseous detectors. With timing precision at the level of tens of picoseconds, this technology is particularly suited for time-of-flight (TOF) applications in particle physics and potentially medical imaging. Beam tests with a 150~GeV/$c$ muon beam have been carried out on a large-area (10~$\times$~10~cm$^{2}$) prototype equipped with a cesium iodide (CsI) photocathode. Using an oscilloscope-based single-channel readout, timing measurements on two individual pads of the detector have yielded resolutions of $\approx$ 48 ps and $\approx$ 52 ps under different biasing conditions respectively.
\end{abstract}

\begin{keywords}
Gaseous detectors \sep Fast timing detectors \sep Photocathodes 
\end{keywords}

\maketitle 
    
\section{Introduction}
Interest in developing technology for precise charged particle timing at high rates has grown due to the potential for pileup-induced backgrounds at the High Luminosity LHC (HL-LHC)~\cite{White:2013taa}. Photodetectors and charged particle detectors with time resolutions in the sub-nanosecond range play a crucial role in both high-energy physics and medical imaging, driving advancements in these fields. In particle physics, the time-of-flight (TOF) technique is used in particle identification through determining the mass of particles with a known momentum. The recent developments in this area has been thoroughly examined in Ref.~\cite{PID}. The $\upmu$RWELL-PICOSEC detector technology was measured the timing performance of 23 ps and 37 ps for a single-channel $\upmu$RWELL-PICOSEC detector with a cesium iodide (CsI) and diamond like carbon (DLC) photocathode, respectively through utilizing a 150~GeV/c muon beam~\cite{akash}.
\par
This manuscript is organized as follows: Section~\ref{sec:concept} introduces the $\upmu$RWELL-PICOSEC detection concept, followed by Section~\ref{sec:assembly} on the design and assembly of the $10 \times 10\ \mathrm{cm}^{2}$ prototype. Section~\ref{sec:electronics} describes the experimental setup and readout electronics, covering both single- and multi-channel configurations. The timing analysis methodology are given in Section~\ref{sec:timming}, with results presented in Section~\ref{sec:results}. Section~\ref{sec:conclusion} concludes with a summary of the findings and outlook for further performance optimization.

\section{Detection concept of $\upmu$RWELL-PICOSEC}
\label{sec:concept}   
\begin{figure}[!ht]   
  \centering
    \includegraphics[width=0.51\textwidth, angle=0]{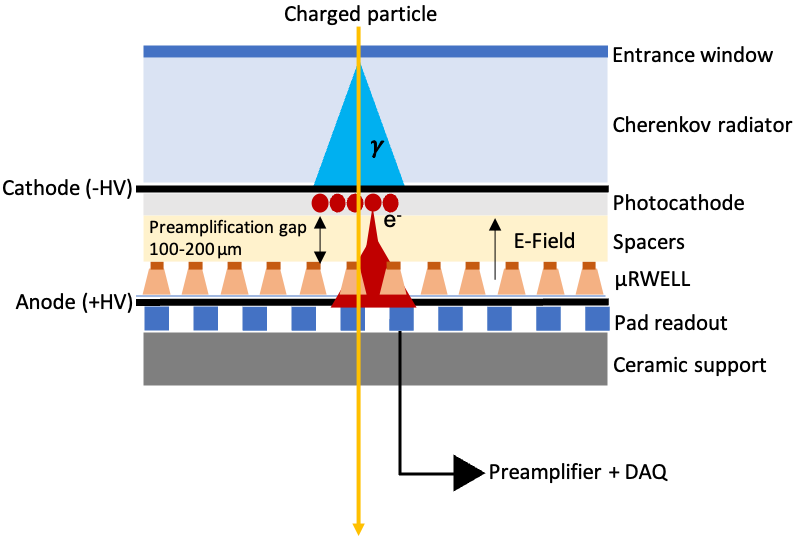}
    \caption{\label{fig:principle}  
    Schematic cross-section of the $\upmu$RWELL-PICOSEC detector. The Cerenkov radiator, photocathode, narrow pre-amplification gap (defined by spacers), and $\upmu$RWELL amplification stage are shown, together with the electric field configuration. The detector is operated with a Ne:C$_2$H$_6$:CF$_4$ (80:10:10) gas mixture at ambient pressure.}
\end{figure}

The $\upmu$RWELL-PICOSEC detector introduces a novel architecture by replacing the conventional 3~mm ionization gap and cathode plane of a standard $\upmu$RWELL detector with a compact photon conversion and amplification scheme. As illustrated in Fig.~\ref{fig:principle}, this design incorporates a Cerenkov radiator positioned above a photocathode. When a high-energy charged particle traverses the radiator, it emits Cerenkov photons ($\gamma$), which subsequently strike the photocathode, releasing photoelectrons. These photoelectrons enter a narrow gas gap, on the order of 100--200~$\upmu$m, defined by precision spacers. Within this region, the photoelectrons  inducing the ionization electrons under the influence of an electric field. The resulting charge is then multiplied further withing the $\upmu$RWELL amplification stage~\cite{Bencivenni_2015}. The detector is operated with a Ne:C$_2$H$_6$:CF$_4$ gas mixture in the ratio 80:10:10 at ambient pressure, providing efficient charge transport and fast signal formation. This arrangement allows for precise timing capabilities because the initial photoelectron generation is localized at the photocathode and the subsequent amplification occurs in a well-shaped geometry. Fig.~\ref{fig:principle} delineates each components of the detector and highlights the electric field configuration essential for efficient charge transport and avalanche multiplication.

\section{Development and assembly of $10 \times 10\ \mathrm{cm}^{2}$ $\upmu$RWELL-PICOSEC detector}
\label{sec:assembly}

A $\upmu$RWELL-PICOSEC prototype with an active area of 10 $\times$ 10 cm$^{2}$ has been constructed and tested as part of ongoing efforts to scale the $\upmu$RWELL-PICOSEC technology. This prototype includes a 100-pad readout layer, with each pad measuring 1~$\times$~1~cm$^{2}$, to extend the fast timing capabilities observed in earlier single channel prototypes. The amplification geometry of $\upmu$RWELL-PICOSEC PCB has been constructed from 50~$\upmu$m thick Kapton dielectric, holes with inner and outer diameter of 80~$\upmu$m and 100~$\upmu$m respectively at a pitch of 120~$\upmu$m.

\begin{figure}[!ht]
    \centering
    \includegraphics[width=0.49\textwidth]{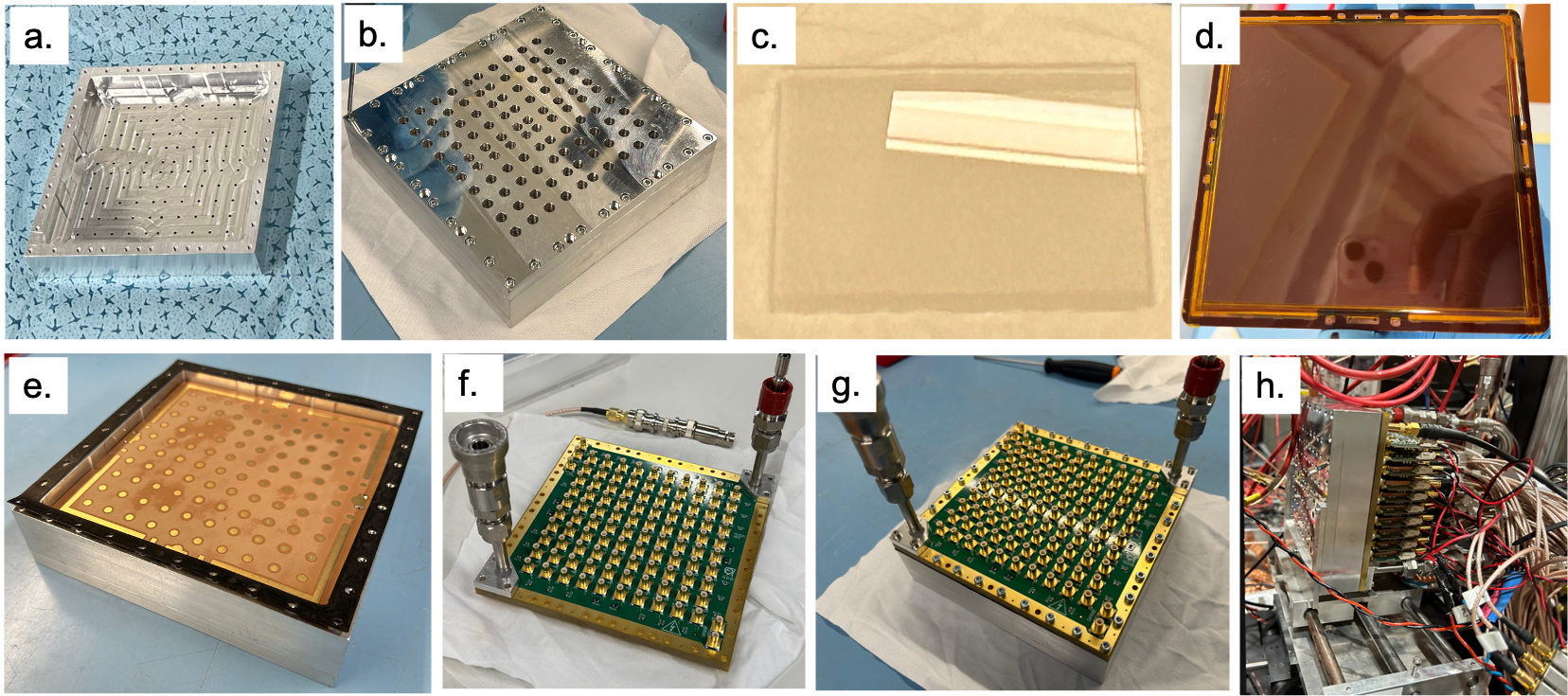} 
    \caption{Photograph showing various components of the $\upmu$RWELL-PICOSEC detector during the assembly process. (a),(b) Aluminum housing (c) CsI photocathode (d) $10 \times 10\ \mathrm{cm}^{2}$ $\upmu$RWELL-PICOSEC PCB (e)Assembled $\upmu$RWELL-PICOSEC PCB with cesium iodide photocathode  (f) Outer board panel (Readout and HV biasing)(g) Fully assembled $\upmu$RWELL-PICOSEC detector (h) Assembled prototype within aluminum casing, with ten 10-channel preamplifier boards connected to the 100 pads on the back of the detector.}
    \label{fig:assembly}
\end{figure}

During assembly, the process begins by positioning the CsI photocathode (c) inside the aluminum housing (a) and (b). A spacer is then inserted to ensure a uniform pre-amplification gap, followed by the placement of the $\upmu$RWELL PCB (d). The assembly is subsequently sealed with the outer board panel (f), which provides both readout and HV biasing. In Fig.~\ref{fig:assembly} are photographs of the various components during assembly, where (a) and (b) depict the aluminum housing, (c) the CsI photocathode, (d) the $\upmu$RWELL PCB, (e) the $\upmu$RWELL PCB positioned on top of the CsI photocathode inside the aluminum casing, (f) the outer board panel for readout and HV biasing, (g) the fully assembled $\upmu$RWELL-PICOSEC detector within its aluminum casing, and (h) the completed prototype with ten 10-channel preamplifier boards connected to the 100 pads on the back of the detector.

\section{Experimental setup and readout electronics of $\upmu$RWELL-PICOSEC detector}
\label{sec:electronics}

\subsection{Experimental setup for timing studies}
\begin{figure}[!ht]
  \centering
    \includegraphics[width=0.5\textwidth, angle=0]{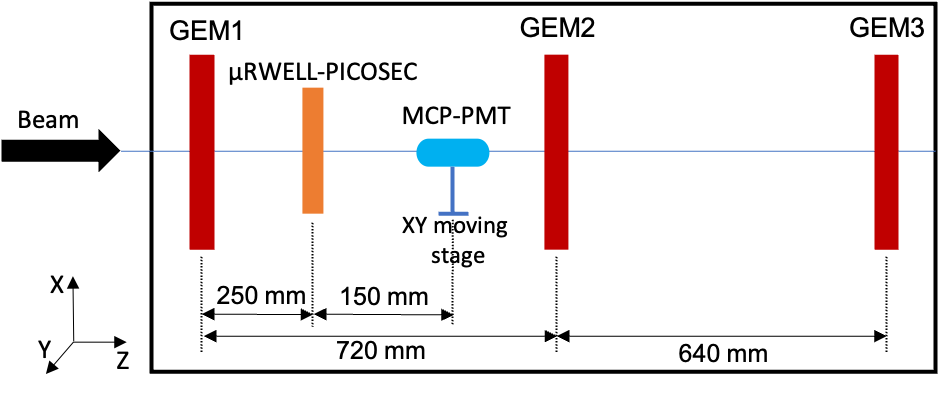}
    \caption{\label{fig:telescope}  A schematic of a telescope configuration with $\upmu$RWELL-PICOSEC detector (in orange), MCP PMT(in blue) for timing reference, and GEM detectors for precise tracking (in red)}
\end{figure}

Beam tests were conducted at the CERN SPS H4 beamline using a 150 GeV/c muon beam. The setup consisted of a compact beam telescope, assembled for precise tracking, trigger generation, and timing reference. The telescope included three 10 cm $\times$ 10 cm triple-GEM detectors for tracking and a MCP-PMT (R3809U-50 Hamamatsu)\cite{MCPPMT} to provide a precise time reference. The MCP-PMT was mounted on an XY moving stage, enabling precise scanning of the 100 pads of the $\upmu$RWELL-PICOSEC detector. A schematic view of the telescope configuration is shown in Fig.~\ref{fig:telescope}. The triple-GEM detectors with a spatial resolution of 70~$\upmu$m, allowed accurate reconstruction of the muon trajectories through the telescope. The MCP-PMT, positioned close to the $\upmu$RWELL-PICOSEC, was used to generate timing triggers for the data acquisition (DAQ) system and to serve as a reference for evaluating the time resolution of the $\upmu$RWELL-PICOSEC detector. The GEM trackers were read out with APV25\cite{APV25}-based SRS electronics\cite{SRS2011} for efficient tracking data acquisition. The compact and modular design of the setup ensured stable beam alignment and consistent test conditions

\subsection{Single-channel readout using oscilloscope}

\begin{figure}[!ht]
  \centering
    \includegraphics[width=0.48\textwidth, angle=0]{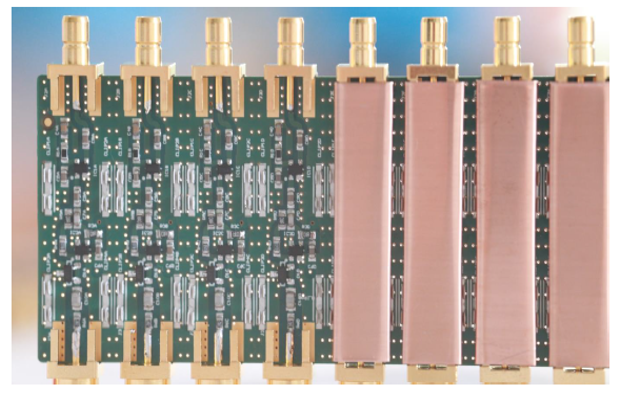}
    \caption{\label{fig:preamp} 
    Ten 10-channel preamplifier boards are used in the $\upmu$RWELL-PICOSEC prototype, 
    each serving ten pads and together providing amplification for all 100 pads. 
    The boards form the common front-end stage for both oscilloscope and SAMPIC readout systems.}
\end{figure}

Ten multi-channel preamplifier boards each with 10 channels serve as the detector front end as shown in Fig.~\ref{fig:preamp}. The amplifier is built around a customized pulse amplifier chip \cite{preamp}, originally developed for CVD diamond detectors and later optimized for PICOSEC applications. Each channel has built-in discharge protection up to 350~V, a bandwidth of 650~MHz, a gain of 38~dB, and a power consumption of 75~mW.

For the oscilloscope-based readout configuration, the signal from a single pad of the $\upmu$RWELL-PICOSEC detector is routed through one channel of this preamplifier before digitization. The amplified signal, together with the MCP-PMT reference signal, recorded using a LECROY WR8104 oscilloscope \cite{oscillo}, operated at an analogue bandwidth of 1.0~GHz and a sampling rate of 10~GSamples/s. This configuration ensured high-precision timing measurements, although it is not scalable to a large number of detector pads.

The digitized signals from the oscilloscopes have been used to analyze the timing and amplitude of the signals, allowing for precise time resolution measurements. However, the use of 100 oscilloscope channels has been found to be highly impractical when scaling to detectors with multiple channels, as the process becomes complex and inefficient.

\subsection{Multichannel readout using SAMPIC}

To overcome the challenges posed by oscilloscopes in large-scale readout, a multi-channel readout system based on the SAMPIC digitizer~\cite{SAMPIC} has been used for the $\upmu$RWELL-PICOSEC detectors. This system enables efficient, high-speed data acquisition for detectors with numerous channels, such as the $\upmu$RWELL-PICOSEC modules.

In this configuration, the same ten 10-channel preamplifier boards (Fig.~\ref{fig:preamp}) have been employed as the amplification stage, routing the signals from all 100 pads to the SAMPIC digitizer. A 64-channel version of the SAMPIC has been successfully tested in recent beam campaigns, and the system has since been expanded to 128 channels to accommodate the full readout of the detector.The SAMPIC system, with a maximum sampling frequency of 8.5~GS/s, has already been validated for timing studies with large-area $10 \times 10\ \mathrm{cm}^{2}$ MM-PICOSEC detectors~\cite{LISOWSKAMM}.

The SAMPIC-based readout effectively addresses the scalability issues of oscilloscopes by managing large data volumes across multiple channels. Its feasibility has been validated through the successful readout of all 100 multi-pad $\upmu$RWELL-PICOSEC channels, with each channel independently digitized to ensure precise timing synchronization and enhanced performance through superior noise filtering and signal processing.

\section{Methodology of timing analysis}
\label{sec:timming}

To assess the timing precision of the $\upmu$RWELL-PICOSEC detector, a reference detector with superior timing capabilities is necessary. In this study, an MCP-PMT (R3809U-50 Hamamatsu), featuring a uniform response over an 11~mm diameter area, has been employed as the timing reference for the $\upmu$RWELL-PICOSEC large prototype. 
The MCP-PMT was aligned with the center of the pad under study, and by translating it across the $10 \times 10\ \mathrm{cm}^{2}$ detector plane using an XY moving stage, the entire active area of the $\upmu$RWELL-PICOSEC has been systematically scanned from one pad center to another.

For the $\upmu$RWELL-PICOSEC, the signal from pad~\#45 was recorded using the oscilloscope-based DAQ system described in Section~\ref{sec:electronics}. These digitized waveforms were then processed by fitting sigmoid functions~\eqref{eq:sigmoid} to the leading edges of the electron peak, which correspond to the charge pulses generated on the anode by avalanche multiplication of the photoelectron cloud. The signal positions were identified at the 20\% Constant Fraction (CF) as described in~\cite{timeanl}.

\begin{equation}
V(t) = \frac{P_{0}}{1+exp(-P_{2}\times (t-P_{1}))} + P_{3}
    \label{eq:sigmoid}
\end{equation}
where $P_{0}$ and $P_{3}$ represent the maximum and minimum values of the signal, respectively; $P_{1}$ corresponds to the inflection point, where the slope of the curve changes sign; and $P_{2}$ characterizes the steepness of the sigmoid change, which is directly related to the signal rise time. 

Timing stamps from the $\upmu$RWELL-PICOSEC and the MCP-PMT corresponding to a 20\% CF have been calculated as given below:

\begin{equation}
t = P_{1}- \frac{1}{P_{2}} \log \left[\frac{P_{0}}{0.2\times V_{\mathrm{max}} - P_{3}} - 1\right]
    \label{eq:cftime}
\end{equation}
where $V_{\mathrm{max}}$ is the electron-peak amplitude.

The signal arrival time (SAT) has been calculated as the difference between the timing marks from the $\upmu$RWELL-PICOSEC detector and the MCP-PMT. The time resolution of the $\upmu$RWELL-PICOSEC detector has been computed by evaluating the standard deviation of the SAT distribution. Notably, the MCP-PMT’s contribution has not been subtracted from the time resolution of the $\upmu$RWELL-PICOSEC detector for any of the measurements presented in this work. The detailed SAT distributions and the corresponding timing resolution results are presented and discussed in Section~\ref{sec:results}.

\begin{figure}[!ht]
    \centering
    \includegraphics[width=0.5\textwidth]{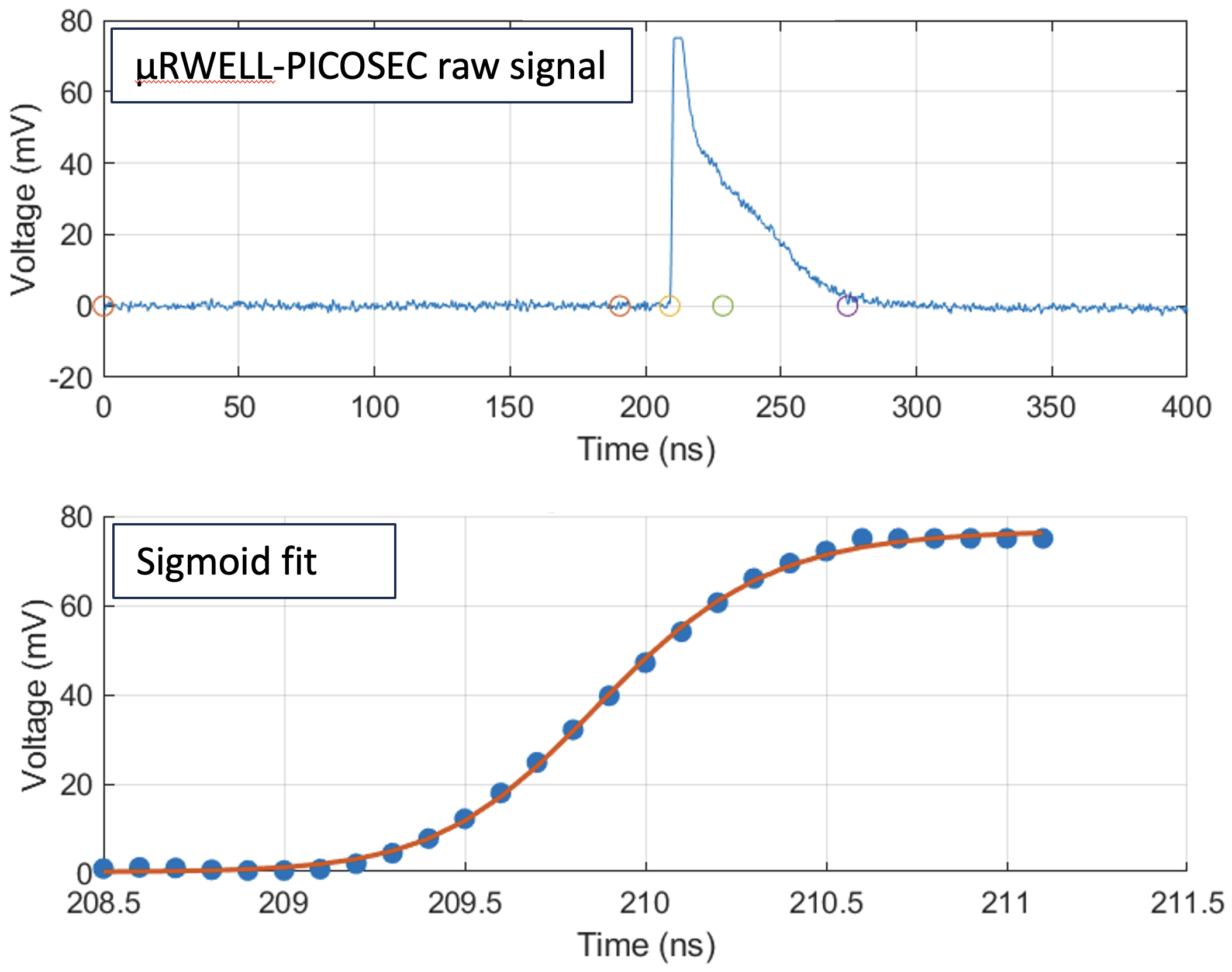} 
    \caption{\label{fig:raw_signal_and_fit} Top: Raw signal from the $\upmu$RWELL-PICOSEC detector acquired with the oscilloscope-based DAQ. Bottom: Sigmoid fit to the leading edge of the electron peak used to calculate the signal arrival time.}
\end{figure}

The sigmoid fit applied to the oscilloscope-recorded signal data from the $\upmu$RWELL-PICOSEC detector, shown in Fig.~\ref{fig:raw_signal_and_fit}, provides an essential tool for determining the precise timing position at the 20\% CF level. The top panel of the figure illustrates the raw signal, while the bottom panel presents the sigmoid function fit to the leading edge of the electron peak, used to extract the signal timing accurately.

\section{Results}
\label{sec:results}

A detailed study of the timing response of the $\upmu$RWELL-PICOSEC detector has been carried out by varying the cathode high voltage while keeping the bias voltage on the $\upmu$RWELL anode layer fixed. The measurements were performed on pad~\#45 of a detector featuring a 170~$\upmu$m drift gap and CsI photocathode. For this study, the signal from pad~\#45 was recorded using the oscilloscope-based DAQ system described in Section~\ref{sec:electronics}, ensuring high-precision waveform capture for subsequent timing analysis. The choice of pad~\#45 was random, as individual pads were independently connected to the oscilloscope during the measurement campaign.  

\begin{figure}[!ht]
    \centering
    \includegraphics[width=0.48\textwidth]{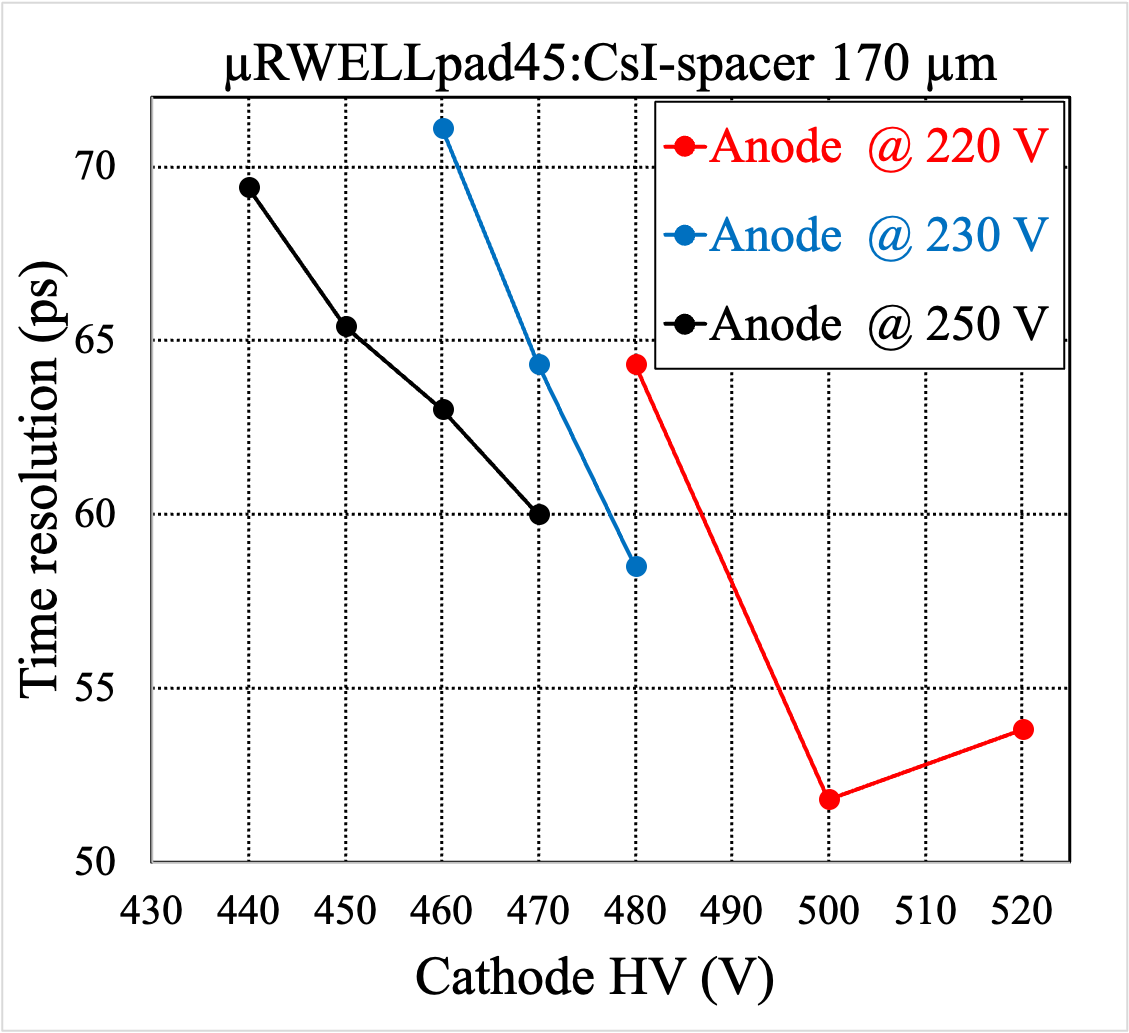}
    \caption{Time resolution of pad~\#45 as a function of cathode high voltage (HV) for a $\upmu$RWELL-PICOSEC detector with a 170~$\upmu$m drift gap and CsI photocathode. Measurements were carried out for three different $\upmu$RWELL anode layer bias settings: 220~V (red), 230~V (blue), and 250~V (black). Data were acquired using the oscilloscope-based DAQ system.}
    \label{fig:hvscan}
\end{figure}

As shown in Fig.~\ref{fig:hvscan}, the time resolution improves with increasing cathode HV for all three settings of the $\upmu$RWELL anode layer: 220~V, 230~V, and 250~V. This improvement is attributed to enhanced electron transport and preamplification in the drift region. The best resolution of approximately 52~ps was achieved for a 220~V bias on the $\upmu$RWELL anode layer and a cathode voltage of 500~V. A slight degradation in resolution was observed beyond this point, which may be related to space-charge effects or signal saturation.

\begin{figure}[!ht]
    \centering
    \includegraphics[width=0.5\textwidth]{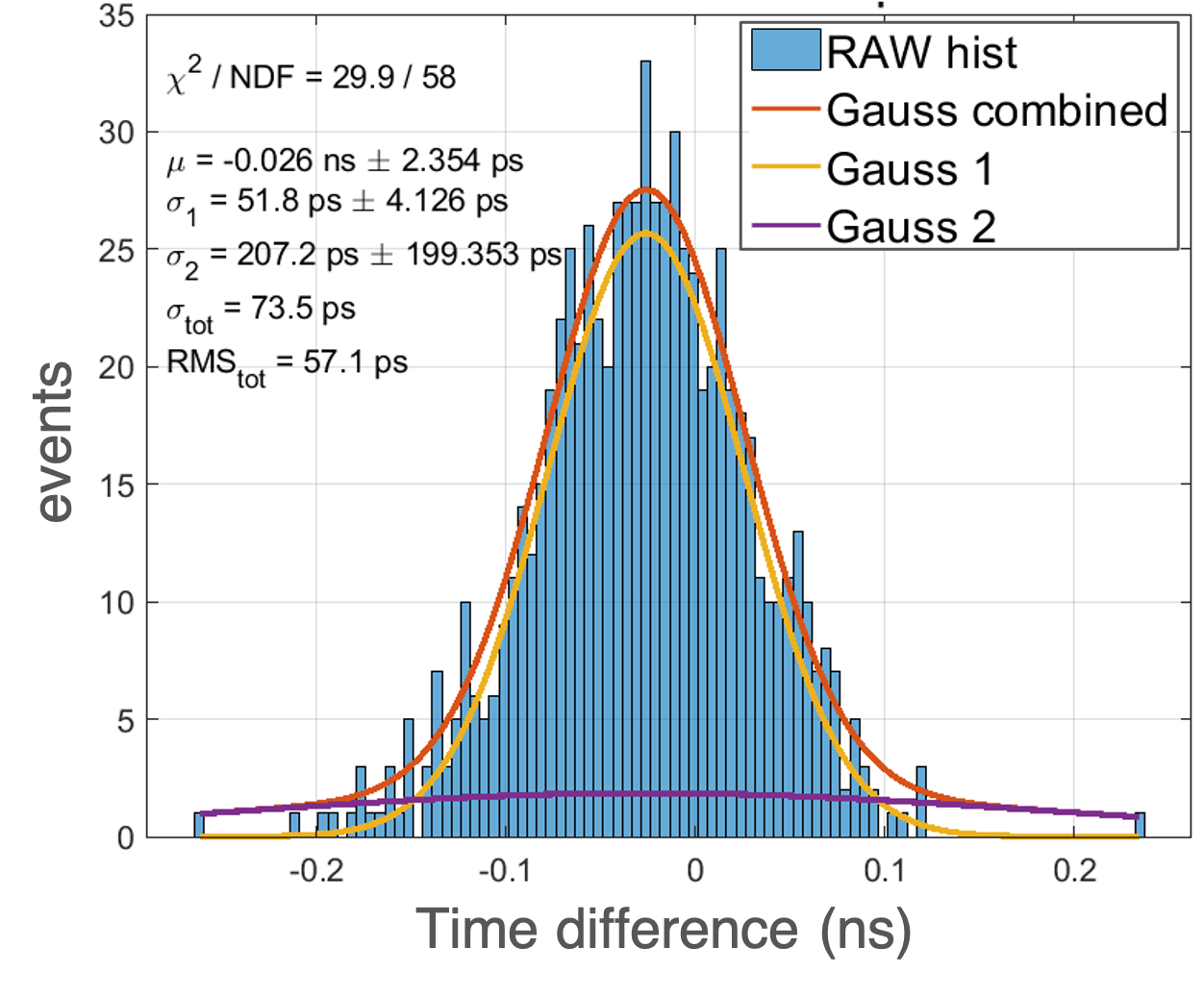}
    \caption{Signal arrival time (SAT) distribution for pad~\#45 at the optimal operating point (cathode HV = 500~V, $\upmu$RWELL anode layer bias = 220~V). The distribution is fitted with a double-Gaussian function, where the narrower Gaussian width corresponds to the intrinsic timing resolution. Data were obtained using the oscilloscope-based DAQ system.}
    \label{fig:sat_distribution}
\end{figure}

The optimal operating point from the HV scan (cathode HV = 500~V, $\upmu$RWELL anode layer bias = 220~V) was further examined using the signal arrival time (SAT) distribution, as defined in Section~\ref{sec:timming}. Figure~\ref{fig:sat_distribution} presents the corresponding distribution of time differences between the $\upmu$RWELL-PICOSEC detector and the MCP-PMT reference. The histogram was fitted with a double-Gaussian function to account for both the core timing response and non-Gaussian tails. The narrower Gaussian width ($\sigma_{1}$) of $51.8 \pm 4.1$~ps represents the intrinsic timing resolution, while the broader component describes residual jitter and other contributions. The total combined width ($\sigma_{\mathrm{tot}}$) was found to be $73.5$~ps, with an overall $\mathrm{RMS}_{\mathrm{tot}}$ of $57.1$~ps.

\begin{figure}[!ht]
    \centering
    \includegraphics[width=0.5\textwidth]{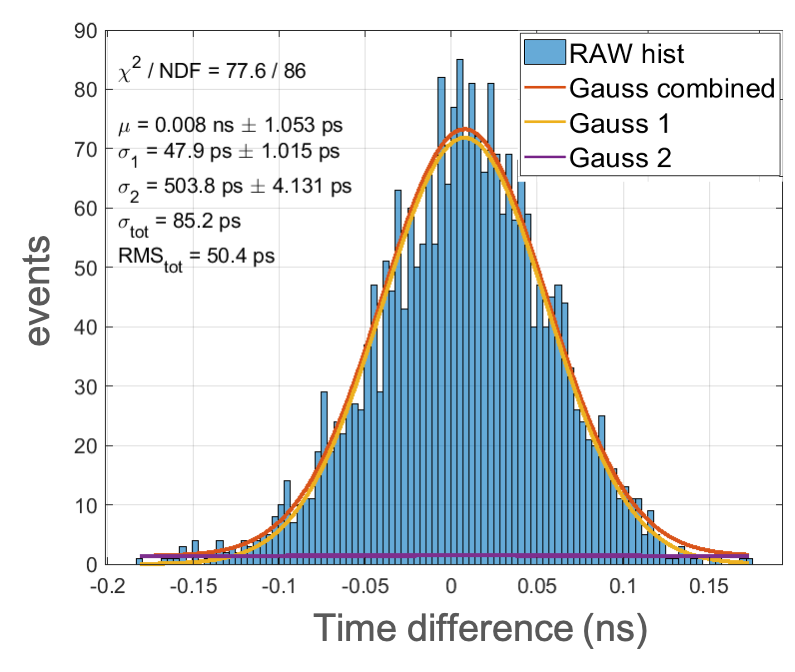}
    \caption{Signal arrival time (SAT) distribution for pad~\#28 at operating conditions of cathode HV = 465~V and $\upmu$RWELL anode layer bias = 250~V. The distribution is fitted with a double-Gaussian function. The narrower Gaussian width of $47.9 \pm 1.0$~ps corresponds to the intrinsic timing resolution. Data were obtained using the oscilloscope-based DAQ system.}
    \label{fig:sat_pad28}
\end{figure}

In addition to pad~\#45, timing studies were also performed on pad~\#28, which—like pad~\#45—was chosen arbitrarily and connected to the oscilloscope for data acquisition. These measurements were carried out under different operating voltages (cathode HV = 465~V, $\upmu$RWELL anode layer bias = 250~V). The corresponding SAT distribution, shown in Fig.~\ref{fig:sat_pad28}, yields an intrinsic timing resolution of $47.9 \pm 1.0$~ps, with a total width of $85.2$~ps and an overall $\mathrm{RMS}_{\mathrm{tot}}$ of $50.4$~ps.

\begin{figure}[!ht]
    \centering
    \includegraphics[width=0.48\textwidth]{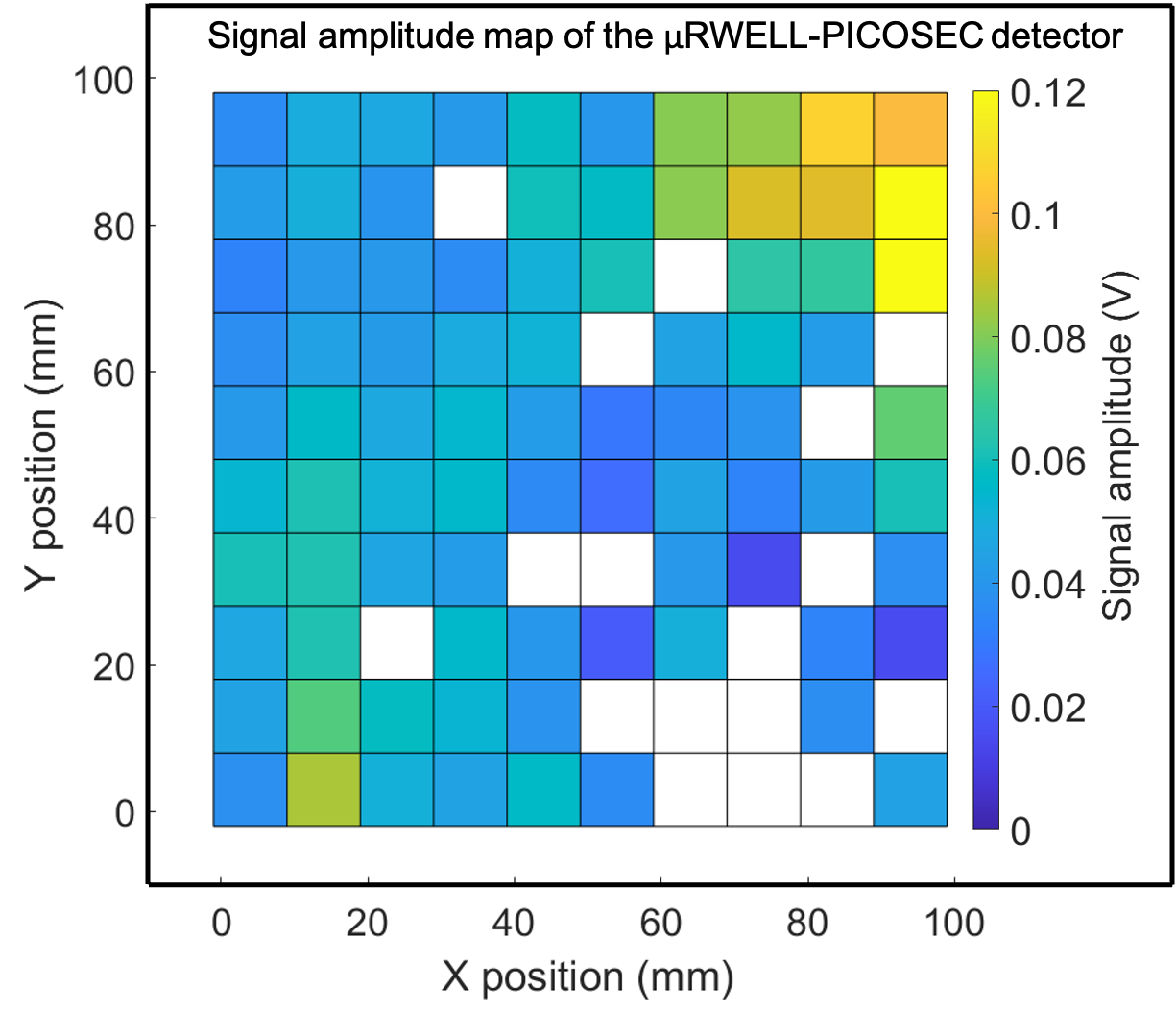}
   \caption{Signal amplitude map of the $10 \times 10\ \mathrm{cm}^{2}$ $\upmu$RWELL-PICOSEC detector from a position scan. The detector comprises 100 readout pads ($1 \times 1\ \mathrm{cm}^{2}$ each). Axes correspond to the detector dimensions in millimeters. White spots correspond to pads whose readout channels were inactive during this measurement.}
    \label{fig:amp_map}
\end{figure}

Finally, a position scan over the full $10 \times 10\ \mathrm{cm}^{2}$ detector plane was performed to study the uniformity of the signal response. The 2D map in Fig.~\ref{fig:amp_map} shows the average signal amplitude (V) recorded by each pad during the scan. Several pads appear blank because their readout channels were inactive during this measurement. Across the active area, significant non-uniformity of the signal amplitude was observed. This variation is suspected to arise from the poor quality of the CsI photocathode used in this prototype, as well as potential flatness issues of the $\upmu$RWELL PCB surface.

\section{Conclusion}
\label{sec:conclusion}
The large-area $10 \times 10\ \mathrm{cm}^{2}$ $\upmu$RWELL-PICOSEC prototype has been successfully developed, assembled, and tested with a 150~GeV/$c$ muon beam at the CERN SPS. The detector, which combines a CsI photocathode, a narrow pre-amplification gap, and a $\upmu$RWELL amplification stage, has demonstrated precise timing capabilities when studied at the pad level. A timing resolution of $51.8 \pm 4.1$~ps has been obtained for pad~\#45 at a cathode HV of 500~V and an anode bias of 220~V, while pad~\#28 has yielded a resolution of $47.9 \pm 1.0$~ps at a cathode HV of 465~V and an anode bias of 250~V. The corresponding SAT distributions confirmed the reliability of the timing methodology in both cases.

Although the timing performance of the large prototype remains more than twice worse than that achieved with earlier small-scale prototypes of similar design~\cite{akash}, this limitation is primarily attributed to the quality of the CsI photocathode and the non-flatness of the PCB. Both issues are expected to be addressed in future iterations, and a timing resolution better than 20~ps is anticipated across the full active area. These results demonstrate the scalability of the $\upmu$RWELL-PICOSEC concept and its potential for large-area, high-precision timing applications in high-energy physics experiments.

\section{Acknowledgement}
The research described in this article was conducted under the Laboratory Directed Research and Development (LDRD) Program at Thomas Jefferson National Accelerator Facility for the U.S. Department of Energy under contract DE-AC05-06OR23177.

\bibliographystyle{unsrt}
\bibliography{refrences.bib}
\end{document}